\documentclass[a4paper]{article}

\usepackage{amsmath,amssymb,graphicx}
\usepackage{subfigure}

\usepackage{lscape}
\usepackage{mathrsfs}

\usepackage{algorithmic}
\usepackage[ruled]{algorithm}

\usepackage{multirow}
\usepackage{verbatim}

\usepackage{color}

\def\ds{\displaystyle}

\newtheorem{lemma}{Lemma}
\newtheorem{theorem}{Theorem}
\newtheorem{proposition}{Proposition}
\newtheorem{corollary}{Corollary}

\def\ba{\begin{array}}
\def\ea{\end{array}}
\def\be{\begin{equation}}
\def\ee{\end{equation}}
\def\bt{\begin{theorem}}
\def\et{\end{theorem}}
\def\bp{\begin{proposition}}
\def\ep{\end{proposition}}
\def\bc{\begin{corollary}}
\def\ec{\end{corollary}}
\def\bo{\begin{proof}}
\def\eo{\end{proof}}
\def\bx{\begin{example}}
\def\ex{\end{example}}
\def\br{\begin{remark}}
\def\er{\end{remark}}
\def\bl{\begin{lemma}}
\def\el{\end{lemma}}

\newbox\bigstrutbox
\setbox\bigstrutbox=\hbox{\vrule height12.5pt depth5pt width0pt}
\def\bigstrut{\relax\ifmmode\copy\bigstrutbox\else\unhcopy\bigstrutbox\fi}

\newbox\Bigstrutbox
\setbox\Bigstrutbox=\hbox{\vrule height17.5pt depth5pt width0pt}
\def\Bigstrut{\relax\ifmmode\copy\Bigstrutbox\else\unhcopy\Bigstrutbox\fi}

\usepackage{amsfonts}
\usepackage{epsfig}
\usepackage{amsmath}
\usepackage{graphicx}
\usepackage{amsmath}
\usepackage{amscd}
\usepackage{stmaryrd}
\usepackage{algorithm}
\usepackage{arydshln}
\usepackage{algorithmic}
\usepackage{youngtab}
\usepackage{young}
\usepackage{geometry}
\usepackage{bbm}
\usepackage{url}

\usepackage[colorlinks,
            linkcolor=blue,
            anchorcolor=blue,
            citecolor=blue,
            urlcolor=black,
            ]{hyperref}

\usepackage{longtable}
\def\u{{\bf u}}
\def\v{{\bf v}}
\newtheorem{remark}{Remark}

\title{Quantum Circulant Preconditioner for Linear System of Equations }

\author{Changpeng Shao\thanks{Key Laboratory of Systems and Control, Academy of Mathematics and Systems Science, Chinese Academy of Sciences, Beijing 100190, P. R. China. (cpshao@amss.ac.cn). }
~and
Hua Xiang\thanks{Corresponding author. School of Mathematics and Statistics, Wuhan University, Wuhan 430072, P. R. China. (hxiang@whu.edu.cn). }
}

\begin{document}
\date{}
\maketitle

\begin{abstract}
We consider the quantum linear solver for $Ax=b$ with the circulant preconditioner $C$.
The main technique is the singular value estimation (SVE) introduced in
\cite[I. Kerenidis and A. Prakash, Quantum recommendation system, in ITCS 2017]{Kerenidis}.
However, some modifications of SVE should be made to solve the preconditioned linear system $C^{-1}Ax=C^{-1}b$.
Moreover, different from the preconditioned linear system considered in \cite[B. D. Clader, B. C. Jacobs, C. R. Sprouse, Preconditioned quantum linear system algorithm, Phys. Rev. Lett., 2013]{CladerJacobsSprouse_PRL13},
the circulant preconditioner is easy to construct and can be directly applied to general dense non-Hermitian cases.
The time complexity depends on the condition numbers of $C$ and $C^{-1}A$, as well as the Frobenius norm $\|A\|_F$.

\end{abstract}




\section{Introduction}

Given $A\in \mathbb{R}^{n\times n}$ and $b \in \mathbb{R}^n$, the linear system of equations $Ax=b$  is a basic problem in scientific computing. The classical methods include direct methods and iterative methods. Gauss elimination with partial pivoting (GEPP) is the generally used direct method. Jacobi method, Gauss-Seidel method, SOR are typical classical iterative methods, and Krylov subspace methods, such as CG, GMRES, BiCGStab, etc. are the modern iterative methods \cite{Demmel_book97,GolubVanLoan_book13,Saad_book2003}. For a general dense matrix, GEPP costs $O(n^3)$; and for a symmetric positive definite matrix,
the CG method  runs with $O(n s \sqrt{\kappa} \log{1/\epsilon})$, where $\kappa$ is the conditioner number, $s$ and  $\epsilon$ stand for the sparsity and the precision respectively.

The first quantum algorithm to solve sparse linear system was proposed by Harrow, Hassidim, and Lloyd \cite{harrow} in 2009, currently known as HHL algorithm for short.
It is exponentially faster than any classical method by calculating the quantum state of the solution, within a running time of $O( (\log n) s^2 \kappa^2 / \epsilon  )$.
Subsequent works have improved the running time of the HHL algorithm to be linear in $\kappa$ \cite{Ambainis_arXiv10} and the precision dependence to $\log(1/\epsilon)$ \cite{ChildsKothariSomma_SICOMP17}.
Ambainis \cite{Ambainis_arXiv10} reduced the condition number dependence from $\kappa^2$ to $\kappa \log^3 \kappa$.
Further work by  Childs, Kothari and Somm \cite{ChildsKothariSomma_SICOMP17} reduced the precision number dependency of the algorithm from $O( \textmd{poly} (1/\epsilon))$ to $O( \textmd{poly} \log(1/\epsilon))$.
The main idea of HHL algorithm is the singular value decomposition (SVD) based on Hamiltonian simulation.
In 2017, Kerenidis and Prakash \cite{Kerenidis} proposed a different method to achieve the SVD, named by the singular value estimation (SVE), with the introduction of a new data structure of quantum information that similar to the idea of qRAM \cite{Giovannetti}.
Later, based on this work, Wossnig, Zhao and Prakash presented the quantum algorithm \cite{WossnigZhaoPrakash_PRL18} to general dense linear systems that takes time $O (\kappa^2 \sqrt{n} \textmd{poly} (\log n) / \epsilon )$, a polynomial speedup for dense matrices.
Wang and Wossnig \cite{Wang-Wossnig18} applied this method for dense Hamiltonian simulation. Some other applications of SVE are given in \cite{Kerenidis} and \cite{Kerenidis1}.
HHL algorithm has wide applications, such as data processing \cite{WiebeBraunLloyd_PRL12}, numerical calculation \cite{MontanaroPallister_PRA16}, artificial intelligence \cite{LloydMohseniRebentrost_arXiv13,RebentrostMohseniLloyd_PRL14}, neural networks \cite{Rebentrost17}, and so on.
It is experimentally demonstrated with parametric down-converted single photons \cite{BarzKassal_SciRep14, CaiPan_PRL13}, liquid nuclear magnetic resonance \cite{PanDu_PRA14},
a scalable superconducting quantum circuit \cite{ZhengPan_PRL17}.

We notice that the condition number $\kappa$ of $A$ plays an important role in the time complexity for both the classical and quantum algorithms.
In order to reduce the dependence on condition number, one important technique is the preconditioning, and we need to solve a preconditioned linear system $M A x = M b$ instead, where the preconditioner $M$ is chosen such that $M \approx A^{-1}$. 
The iterative methods are successful only if there exists an effective preconditioner. For example, the classical CG on the typical 2nd-order elliptic boundary value problems in 3D, using the preconditioner can reduce the conditioner number from $O(n^{2/3})$ to $O(n^{1/3})$, and the time complexity of $O(n^{4/3})$ decreases to $O(n^{7/6})$.
There exists many preconditioning techniques for the classical methods
\cite{Demmel_book97,GolubVanLoan_book13,Saad_book2003}, including AMG, DDM, etc. To the best of our knowledge, there was only one work related to the quantum preconditioning \cite{CladerJacobsSprouse_PRL13}.
To improve the efficiency of quantum linear solver,
Clader et al. \cite{CladerJacobsSprouse_PRL13} chose a sparse approximate inverse (SPAI) preconditioner $M$. It needs a unitary operator to calculate the elements of $MA$. The oracle for the matrix $MA$ can be created by using the original oracle for $A$ with only modest overhead of $O(s^3)$ in run time and $O(s^2)$ in query complexity. Under the sparsity assumption of $MA$, this work improves the complexity of HHL algorithm to $O(s^7\kappa(MA)(\log n)/\epsilon)$.

In this paper, we consider another kind of preconditioner, a circulant preconditioner $C$.
Different from the SPAI used in \cite{CladerJacobsSprouse_PRL13}, the circulant preconditioner is more general \cite{Chan_SISC88}, suitable for the general dense linear systems. Moreover, the circulant preconditioner $C$ contains a simple structure. It can be diagonalized by the Fourier transformation. In quantum computing, the quantum Fourier transformation can be implemented efficiently. Hence in some sense $C$ can be just viewed as a diagonal matrix. Such quantum preconditioner is easy to construct and suitable for quantum implementation. The only difficulty that lies in the construction of circulant preconditioner is that the eigenvalues of $C$ is given by a summation. Direct calculation based on such formula costs at least $O(n)$, which kills the exponential speedup of the quantum algorithm. Therefore we should find another efficient method to obtain them.
The main technique that we will use to solve the preconditioned linear system $C^{-1}Ax=C^{-1}b$ is the SVE. However, the SVE given in \cite{Kerenidis} is not sufficient to our problem here, since $C$ is not Hermitian and should be provided in a quantum state form.
So we need to make some modifications about the SVE method introduced in \cite{Kerenidis}.
Assuming the SVD of $A=\sum \sigma_i|u_i\rangle\langle v_i|$, then the SVE given in \cite{Kerenidis} achieves $\sum \alpha_i|v_i\rangle\mapsto \sum \alpha_i|v_i\rangle|\sigma_i\rangle$. However, it will be more helpful to us if we can achieve
$\sum \alpha_i|v_i\rangle\mapsto \sum \alpha_i|u_i\rangle|\sigma_i\rangle$ or $\sum \alpha_i|u_i\rangle\mapsto \sum \alpha_i|v_i\rangle|\sigma_i\rangle$. This can be achieved by making some modifications about the SVE proposed by \cite{Kerenidis}.
For Hermitian matrix $A$, we know that $|u_i\rangle=|v_i\rangle$. For the non-Hermitian matrix $A$, as introduced in HHL algorithm \cite{harrow},  the SVE in \cite{Kerenidis} works on
$\left(
   \begin{array}{cc}
     0 & A \\
     A^\dag & 0 \\
   \end{array}
 \right)$ instead. 
However, if $A$ is given as quantum information, then it may not be easy to expand $A$ into a Hermitian matrix.
But our modified SVE method does not need such expansion and works well on the original non-Hermitian matrix, and hence it can solve the preconditioned linear system $C^{-1}Ax=C^{-1}b$ more efficiently.

The structure of this paper is as follows. In section 2, we briefly review the basic results of classical circulant preconditioner.
Then in section 3, we will introduce the modified SVE method and apply it to solve the preconditioned linear system.

\section{Circulant Preconditioner}

In this section, we will briefly review some basic known results about circulant preconditioner given by Strang \cite{Strang_SAM86}, Chan \cite{Chan_SISC88} and Tyrtyshnikov \cite{Tyrtyshnikov_SIMAX92}.
A circulant preconditioner is defined by an $n$-by-$n$ circulant matrix
\be
C = (c_{ij})_{n\times n} =
\begin{pmatrix}
c_0     & c_{n-1}   & \cdots    & c_2       & c_1   \\
c_1     & c_0       & c_{n-1}   & \cdots    & c_2   \\
\vdots  & c_1       & c_0       & \ddots    & \vdots    \\
c_{n-2} & \cdots    & \ddots    & \ddots    & c_{n-1}    \\
c_{n-1} & c_{n-2}   & \cdots    & c_1       & c_0       \\
\end{pmatrix},
\ee
where the entry $c_{ij} = c_{(i-j)\mod n}$. Obviously the matrix $C$ is totally determined by its first column. Let $Q$ be the following shift permutation matrix
\[
Q=\left(
  \begin{array}{ccccc}
    0 & 0 & 0 & \cdots & 1 \\
    1 & 0 & 0 & \ddots & \vdots \\
    0 & 1 & \ddots & \ddots & 0 \\
    \vdots & \ddots & \ddots & \ddots & 0 \\
    0 & \cdots & 0 & 1 & 0 \\
  \end{array}
\right) ,
\]
then $C=\sum_{j=0}^{n-1} c_j Q^j$.  In \cite{ZhouWang}, Zhou and Wang applied this decomposition for the Hamiltonian simulation of $C$ and solve the linear system $Cx=b$.

The circulant matrix can be diagonalized by Fourier matrix $F=(\frac{1}{\sqrt{n}}\omega^{jk})_{n\times n}$, where $\omega =  e^{-2\pi i / n}$. That is, there is a diagonal matrix $\Lambda = \text{diag} (\lambda_0, \cdots, \lambda_{n-1})$, which refers to the eigenvalues of $C$, such that
\be
C = F^\dagger \Lambda F.
\ee
More precisely, set $e_0=(1,0,\ldots,0)^\dag$, $e=(1,1,\ldots,1)^\dag$, then $F C e_0 = \Lambda F e_0  = \tfrac{1}{\sqrt{n}} \Lambda e $. Note that $C e_0=(c_0, c_1,\ldots, c_{n-1})^\dag$, and $\Lambda e=( \lambda_0, \lambda_1,\ldots, \lambda_{n-1} )^\dag$, then
\be \label{eigenvalue of circulant matrix}
\lambda_k = \sum_{j = 0}^{n-1} c_j \omega^{jk} .
\ee

Let $U$ is a unitary matrix. Define
\begin{equation*}
\mathcal{M}_U := \{ U^\dag \Lambda_n U | \Lambda_n \text{ is an $n \times n$ diagonal matrix} \}.
\end{equation*}
Then $\mathcal{M}_F$ is the set of all circulant matrices.

The Strang preconditioner is designed for the Toeplitz matrix
$$
T = (t_{ij})_{n\times n}=
\begin{pmatrix}
t_0     & t_{-1}    & \cdots    & t_{2-n}   & t_{1-n}   \\
t_1     & t_0       & \ddots    & \cdots    & t_{2-n}   \\
\vdots  & \ddots      & \ddots    & \ddots    & \vdots    \\
t_{n-2} & \cdots    & \ddots    & \ddots    & t_{-1}    \\
t_{n-1} & t_{n-2}   & \cdots    & t_1       & t_0       \\
\end{pmatrix},
$$
i.e., $t_{ij} = t_{i-j}$, determined by $2n-1$ entries.
The matrix name arises from Toeplitz's work on bilinear forms associated with Laurant series.
A circulant matrix is a special case of Toeplitz matrix with $t_{-k} = t_{n-k}$ for $1 \leq k \leq n-1$.
The Toeplitz linear system $T x = b $ appears in a variety of applications, such as signal processing, control theory, networks, integral equations, etc.
The quantum algorithm to the Toeplitz linear system $T x = b $ has been considered in \cite{wan}.
For such linear system, Strang \cite{Strang_SAM86} proposed a circulant preconditioner $s_F(T)$, which satisfies \cite{Chan_SIMAX89}
\begin{equation*}
s_F (T) = \arg\min_{ C \in \mathcal{M}_F } \| T - C \|_1 .
\end{equation*}
For simplicity, we assume that $n = 2m+1$, while the case $n=2m$ can be treated similarly.
The Strang preconditioner $s_F(T)$ is a circulant matrix defined by its first column $s = [s_0, \cdots, s_{n-1}]^T$, where
\begin{equation*}
s_k=\left\{
      \begin{array}{cl}\vspace{.2cm}
        t_k,        & \hbox{$0 \leq k \leq m$,} \\
        t_{k-n},    & \hbox{$m \le k \leq n-1$.}
      \end{array}
    \right.
\end{equation*}

Chan \cite{Chan_SISC88} proposed the optimal circulant preconditioner for solving Toeplitz systems and extended it for general matrices.
For an arbitrary matrix $A$, it can prove that
\begin{equation} \label{eqn:DefOptimalCircuPre}
c_U (A) := U^\dag \text{diag}(U A U^\dag) U = \arg\min_{W\in \mathcal{M}_U} \|A - W \|_F,
\end{equation}
where $\|\cdot\|_F$ is the Frobenius norm, and $\text{diag}(A)$ is the diagonal matrix defined by the main diagonal entries of $A$. The special case $c_F (A)$ is called the optimal circulant preconditioner \cite{Chan_SISC88}.
Then it is easy to see that
\be \label{optimal circulant preconditioner:general}
c_F( A )=\sum_{j=0}^{n-1} \Big(\frac{1}{n}\sum_{p-q\equiv j(\textmd{mod}~n)}a_{pq}\Big)Q^j .
\ee
Especially, when $A =T $ is a Toeplitz matrix, then the entries of circulant preconditioner are given by
\begin{equation*} 
c_k = \left[ (n-k)t_k + k t_{k-n} \right] / n, \quad  (0\leq k\leq n-1).
\end{equation*}

Tyrtyshnikov \cite{Tyrtyshnikov_SIMAX92} suggested a so-called super optimal circulant preconditioner for arbitrary matrix.
We can prove that \cite{ChanJinYeung_SINUM91,Tyrtyshnikov_SIMAX92}
\begin{equation*}
 t_U (A) := c_U(A A^\dag) [c_U(A^\dag)]^{-1} = \arg\min_{W\in \mathcal{M}_U} \|I_n - W^{-1} A \|_F.
\end{equation*}
The special case $t_F (A)$ is called the super-optimal circulant matrix \cite{Tyrtyshnikov_SIMAX92}, where the construction of $t_F(T)$ needs $O(n\log n)$ operations.

To examine the efficiency of the circulant preconditioner, we are concerned about the spectra of preconditioned matrix $C^{-1} A $, where $C$ is a circulant preconditoner. The analysis for general case is difficult. Numerical tests shows that in most cases, the circulant preconditoner can make the condition number of $C^{-1} A $ small.
However, as for Toeplitz matrix $T_n$ with positive generating function in the Wiener class. The circulant preconditoner $C= s_F(T), c_F(T)$ or $t_F(T)$ introduced above satisfy that for all $\epsilon \ge 0$, there exist integers $M$ and $N$, such that for all $n>N$, the matrix $ C^{-1} T -  I_n$ has at most $M$ eigenvalues in absolute value larger than $\epsilon$ \cite{ChanJin_SIAMbook07}. That is, for a large $n$, the spectrum of the preconditioned matrix $C^{-1} T $ is clustered around 1.  We can also prove that the smallest eigenvalue of preconditioned matrix $C^{-1} T $ is uniformly bounded away from the origin. It follows that we can expect the superlinear convergence of preconditioned CG method.

Although the circulant preconditoner $C=c_F(A)$ given in \eqref{eqn:DefOptimalCircuPre} or (\ref{optimal circulant preconditioner:general}) has an explicit formula, to compute all the entries of $C$ will take about $O(n^2)$ in classical computer and at least $O(n)$ in quantum computer. So direct computation of $C$ will bring no benefits in solving the linear system $C^{-1}Ax=C^{-1}b$. The SVE technique only requires the quantum state of $C$, and shows great advantages to solve the circulant preconditoned linear system as we will discuss in the next section.

\section{Preconditioned linear system}

In this section, we consider the preconditioning technique on solving the linear system $Ax=b$. The circulant preconditioner $C$ of this linear system can be constructed, for example, by \eqref{eqn:DefOptimalCircuPre}.
Then the preconditioned linear system reads
\begin{equation} \label{eqn:precondLin}
C^{-1} Ax=C^{-1} b.
\end{equation}

The method we will use to solve the linear system (\ref{eqn:precondLin}) is based on singular value estimation (SVE) introduced in \cite{Kerenidis}.
In subsection \ref{Singular value estimation}, we first introduced the the SVE technique with some modifications.
Then in subsection \ref{Preconditioned linear solver}, we show how to solve (\ref{eqn:precondLin}) based on the modified SVE.

\subsection{Singular value estimation}
\label{Singular value estimation}

In \cite{Kerenidis}, Kerenidis and Prakash introduced a data structure to store matrices in quantum computer efficiently. Based on this data structure, a fast quantum algorithm to the SVE can be obtained. With this SVE technique, we can perform various scientific calculations with quantum computer, such as implementation of dense Hamiltonian simulation \cite{Wang-Wossnig18}, solving dense linear system \cite{WossnigZhaoPrakash_PRL18}, as well as some other applications based on singular value decomposition \cite{Kerenidis1}.

In their original paper \cite{Kerenidis}, the authors used the rows of the given matrix $A$.
Taking into account the preconditioning, here we slightly modify it and use the columns instead.
Let $A=(A_{ij})_{n\times n}$ be a $n\times n$ matrix. For any $0\leq j\leq n-1$, denote $\|A_j\|$ and $|A_j\rangle=\frac{1}{\|A_j\|}\sum_{i=0}^{n-1}A_{ij}|i\rangle$ as the 2-norm and the quantum state of $j$-th column of $A$,
and also define $\|A\|_F=\sqrt{\sum_j\|A_j\|^2}$ as the Frobenius norm of $A$ and $|A_F\rangle=\frac{1}{\|A\|_F}\sum_{j=0}^{n-1}\|A_j\||j\rangle$.
With the similar analysis as \cite{Kerenidis}, the quantum computer can perform the following mappings in $O(\textmd{poly} (\log n))$ time:
\be\ba{lll} \vspace{.2cm}\label{data}
U_\mathcal{M}:|0\rangle|j\rangle&\mapsto&|A_j\rangle|j\rangle=\ds\frac{1}{\|A_j\|}\sum_{i=0}^{n-1}A_{ij}|i,j\rangle, \\
U_\mathcal{N}:|i\rangle|0\rangle&\mapsto&|i\rangle|A_F\rangle=\ds\frac{1}{\|A\|_F}\sum_{j=0}^{n-1}\|A_j\||i,j\rangle.
\ea\ee

Define two degenerate operators $\mathcal{M}$ and $\mathcal{N}$ as
\[
\mathcal{M}:|j\rangle\mapsto|A_j\rangle|j\rangle, \hspace{.5cm}
\mathcal{N}:|i\rangle\mapsto|i\rangle|A_F\rangle.
\]
That is,
\[
\mathcal{M}=\sum_{j=0}^{n-1}|A_j\rangle|j\rangle\langle j|, \hspace{.5cm}
\mathcal{N}=\sum_{i=0}^{n-1}|i\rangle|A_F\rangle\langle i|.
\]
Then we can verify that
\[
\mathcal{N}^\dagger\mathcal{M}=\sum_{i,j=0}^{n-1}|i\rangle\langle i|A_j\rangle\langle A_F|j\rangle \langle j|
=\sum_{i,j=0}^{n-1}\frac{A_{ij}}{\|A\|_F} |i\rangle\langle j|=\frac{A}{\|A\|_F}.
\]
It is also easy to check that $\mathcal{M}^\dagger\mathcal{M}=\mathcal{N}^\dagger\mathcal{N}=I_n$. The following unitary transformation
\[
2\mathcal{M}\mathcal{M}^\dagger-I_{n^2}=2\sum_{j=0}^{n-1}|A_j\rangle|j\rangle\langle A_j|\langle j|-I_{n^2}
=U_\mathcal{M} \left(2\sum_{j=0}^{n-1}|0\rangle|j\rangle\langle 0|\langle j|-I_{n^2}\right) U_\mathcal{M}^\dagger,
\]
can be efficiently implemented in time $O(\textmd{poly} (\log n))$. Similarly, $2\mathcal{N}\mathcal{N}^\dagger-I_{n^2}$ can be efficiently implemented in time $O(\textmd{poly} (\log n) )$ too. Now denote $W=(2\mathcal{N}\mathcal{N}^\dagger-I_{n^2})(2\mathcal{M}\mathcal{M}^\dagger-I_{n^2})$.

Let $A=\sum_{i=0}^{n-1} \sigma_i|u_i\rangle\langle v_i|$ be the singular value decomposition of $A$, then
\[\ba{lll}\vspace{.2cm}
W\mathcal{M}|v_i\rangle &=& (2\mathcal{N}\mathcal{N}^\dagger-I_{n^2})(2\mathcal{M}\mathcal{M}^\dagger-I_{n^2})\mathcal{M}|v_i\rangle \\\vspace{.2cm}
&=& (2\mathcal{N}\mathcal{N}^\dagger-I_{n^2})\mathcal{M}|v_i\rangle \\\vspace{.2cm}
&=& \frac{2}{\|A\|_F}\mathcal{N}A|v_i\rangle-\mathcal{M}|v_i\rangle \\
&=& \frac{2\sigma_i}{\|A\|_F}\mathcal{N}|u_i\rangle-\mathcal{M}|v_i\rangle,
\ea\]
and
\[\ba{lll}\vspace{.2cm}
W\mathcal{N}|u_i\rangle &=& (2\mathcal{N}\mathcal{N}^\dagger-I_{n^2})(2\mathcal{M}\mathcal{M}^\dagger-I_{n^2})\mathcal{N}|u_i\rangle \\\vspace{.2cm}
&=& (2\mathcal{N}\mathcal{N}^\dagger-I_{n^2})(\frac{2}{\|A\|_F}\mathcal{M}\mathcal{A}^\dagger|u_i\rangle-\mathcal{N}|u_i\rangle) \\\vspace{.2cm}
&=& (2\mathcal{N}\mathcal{N}^\dagger-I_{n^2})(\frac{2\sigma_i}{\|A\|_F}\mathcal{M}|v_i\rangle-\mathcal{N}|u_i\rangle) \\\vspace{.2cm}
&=& \frac{4\sigma_i}{\|A\|_F^2}\mathcal{N}A|v_i\rangle-\frac{2\sigma_i}{\|A\|_F}\mathcal{M}|v_i\rangle-\mathcal{N}|u_i\rangle \\
&=& (\frac{4\sigma_i^2}{\|A\|_F^2}-1)\mathcal{N}|u_i\rangle-\frac{2\sigma_i}{\|A\|_F}\mathcal{M}|v_i\rangle.
\ea\]
The subspace $\{\mathcal{M}|v_i\rangle,\mathcal{N}|u_i\rangle\}$ is invariant under $W$. Moreover, $W$ is a rotation in the this space. The matrix representation of $W$ in this space is
\[
W_i=\left(
      \begin{array}{cc}\vspace{.2cm}
        \frac{4\sigma_i^2}{\|A\|_F^2}-1 & \frac{2\sigma_i}{\|A\|_F} \\
        -\frac{2\sigma_i}{\|A\|_F}      & -1 \\
      \end{array}
    \right).
\]
The eigenvalues of $W_i$ are
\[
\frac{2\sigma_i^2}{\|A\|_F^2}-1\pm i \sqrt{1-\Big(\frac{2\sigma_i^2}{\|A\|_F^2}-1\Big)^2}
 \equiv e^{\pm i\theta_i},
\]
where $\theta_i$ satisfies
\[
\cos\theta_i=\langle u_i|\mathcal{N}^\dagger W\mathcal{N}| u_i\rangle=\frac{2\sigma_i^2}{\|A\|_F^2}-1.
\]

We can perform the phase estimation algorithm on $W$ to get the estimates of $\theta_i$.
Then we can compute the singular values of $A$ based on the formula $\sigma_i = \|A\|_F \cos(\theta_i/2)$. This is the main idea of SVE considered in \cite{Kerenidis}.

Note that
\begin{eqnarray*}
(\cos  \tfrac{\theta_i}{2})^{-1} (W_i - e^{\pm i \theta_i} I)
    =  \left(
      \begin{array}{cc}\vspace{.2cm}
       e^{\mp i\tfrac{\theta_i}{2}}   & 1 \\
        -1 & -e^{\pm i\tfrac{\theta_i}{2}}   \\
      \end{array}
    \right) .
\end{eqnarray*}
%
The corresponding eigenvectors of $W$ are
$$\textbf{x}_\pm = e^{\pm i \theta /2} \mathcal{N} | u_i \rangle - \mathcal{M} |v_i \rangle .$$
The vectors $ \mathcal{M} |v_i\rangle $ and $ \mathcal{N} |u_i\rangle $ can be reformulated by $\textbf{x}_\pm$ as follows.
$$ \mathcal{M} |v_i\rangle = ( e^{-i \tfrac{\theta_i}{2} } \textbf{x}_+ - e^{i \tfrac{\theta_i}{2} } \textbf{x}_-) / ( 2 i \sin \tfrac{\theta_i}{2} ) ,
\qquad \mathcal{N} |u_i\rangle = (\textbf{x}_+ - \textbf{x}_-) / ( 2 i \sin \tfrac{\theta_i}{2} ) . $$

Given any state
$  |b\rangle  = \sum_{i=0}^{n-1} \beta_i |v_i \rangle$, we have
$$ U_\mathcal{M} | b\rangle  =  \sum_{i=0}^{n-1} \beta_i \mathcal{M} |v_i \rangle
\equiv \sum_{i=0}^{n-1} \beta_i  \left( e^{-i \tfrac{\theta_i}{2} } m_+ |x_+\rangle - e^{i \tfrac{\theta_i}{2} } m_- |x_-\rangle \right) / ( 2 i \sin \tfrac{\theta_i}{2} ) ,$$
 where $\textbf{x}_\pm = m_\pm |x_\pm\rangle $ and $m_\pm$ are the norms of $\textbf{x}_\pm$.

Using phase estimation algorithm and an oracle for computing $\sigma_i$, that is $\sigma_i = \|A\|_F \cos(\theta_i/2)$, we have
$$  \sum_{i=0}^{n-1} \beta_i  \left( e^{-i \tfrac{\theta_i}{2} } m_+ |x_+\rangle |\theta_i\rangle - e^{i \tfrac{\theta_i}{2} } m_- |x_-\rangle |-\theta_i\rangle  \right) | \sigma_i \rangle  / ( 2 i \sin \tfrac{\theta_i}{2} ) .  $$
Using the phase rotation, the state is transformed into
$$  \sum_{i=0}^{n-1} \beta_i  \left(   m_+ |x_+\rangle |\theta_i\rangle -   m_- |x_-\rangle |-\theta_i\rangle  \right) |\sigma_i \rangle / ( 2 i \sin \tfrac{\theta_i}{2} ) .  $$
Undo the phase estimation algorithm, we then obtain
$$  \sum_{i=0}^{n-1} \beta_i  \left(  m_+ |x_+\rangle  -   m_- |x_-\rangle    \right) |\sigma_i \rangle / ( 2 i \sin \tfrac{\theta_i}{2} )
=  \sum_{i=0}^{n-1} \beta_i  \mathcal{N} |u_i \rangle |\sigma_i \rangle .  $$
Finally, applying $U_\mathcal{N}^{-1}$, we have the state $\sum \beta_i |u_i \rangle |\sigma_i \rangle$. The procedure is summarized in the following lemma.

\bl \label{sve}
Let $A$ be an $n\times n$ matrix with the singular value decomposition $A=\sum_{i=0}^{n-1} \sigma_i|u_i\rangle\langle v_i|$. Then there is a quantum algorithm that runs in $O(\emph{poly} (\log n)/\epsilon)$ and achieves $\sum_i\alpha_i|v_i\rangle|0\rangle \mapsto \sum_i\alpha_i| u_i\rangle|\tilde{\sigma}_i\rangle$, where $|\tilde{\sigma}_i-\sigma_i|\leq \epsilon\|A\|_F$ for all $i$ with probability at least $1-1/\emph{poly}(n).$
\el

Note that
in the paper \cite{Kerenidis}, they achieved $\sum_i\alpha_i|v_i\rangle|0\rangle \mapsto \sum_i\alpha_i| v_i\rangle|\tilde{\sigma}_i\rangle$. However, the result in lemma \ref{sve} is the transformation $\sum_i\alpha_i|v_i\rangle|0\rangle \mapsto \sum_i\alpha_i| u_i\rangle|\tilde{\sigma}_i\rangle$. This procedure is quite suitable to perform matrix multiplication.
Similarly we can efficiently perform the transformation: $\sum_i \alpha_i |u_i\rangle|0\rangle \mapsto \sum_i\alpha_i| v_i\rangle|\tilde{\sigma}_i\rangle$, and such procedure benefits the inverse operation of a matrix.

Usually, when $A$ is non-Hermitian, we need to expand it to a Hermite matrix $\left(
      \begin{array}{cc}
       0        & A \\
       A^\dag   & 0 \\
      \end{array}
    \right)$ and so $|v_i\rangle=|u_i\rangle$.
But for some cases where the matrix is given as quantum information, like the problem considered in this paper, such expansion is not to easy to be realized.
Our method given in lemma \ref{sve} works for non-Hermitian matrix, and does not need such expansion.

\br
Based on the data structure given in \cite{Kerenidis}, similarly we can obtain
\be \label{eqn:stateA}
|A\rangle=\frac{1}{\|A\|_F} \sum_{i,j=0}^{n-1} A_{ij}|i,j\rangle=\frac{1}{\|A\|_F} \sum_{j=0}^{n-1} \|A_j\| |A_j\rangle|j\rangle,
\ee
in time $O(\emph{poly} (\log n) )$.
The SVE in lemma 1 is realized by using $U_\mathcal{M}$ and $U_\mathcal{N}$. However, using the $U_\mathcal{M}$ in \eqref{data} and $|A\rangle$ in \eqref{eqn:stateA}, we can also construct the SVE.
The reason is that
if we apply $U_\mathcal{M}^{-1}$ on $|A\rangle$, then we will get $|A_F\rangle$, equivalently, we obtain $U_\mathcal{N}$. This is the main idea that will be used in our next section.
We just need to focus on the construction of the quantum states of $A$ and its columns.
\er

For the linear system, we can choose $|b\rangle=\sum_i\beta_i|u_i\rangle$. The solution of the linear system $Ax=b$ can be obtained by lemma \ref{sve} in the following way, a similar procedure as HHL algorithm,
\[
\sum_i \beta_i|u_i\rangle |0\rangle \mapsto \sum_i \beta_i|v_i\rangle|\tilde{\sigma}_i\rangle|0\rangle \mapsto
\sum_i \beta_i|v_i\rangle|\tilde{\sigma}_i\rangle\left(Z\tilde{\sigma}_i^{-1}|0\rangle+\sqrt{1-Z^2\tilde{\sigma}_i^{-2}}|1\rangle\right)
\]
for some parameter $Z$. The complexity to get the solution to accuracy $\epsilon$ is about $O(\kappa^2 \textmd{poly} (\log n) \|A\|_F/\epsilon)$. The analysis is the same as HHL algorithm, see \cite{WossnigZhaoPrakash_PRL18}. 

\bl \label{linear-system}
For any matrix $A$ and quantum state $|b\rangle$, the quantum state of $A^{-1}|b\rangle$ to the accuracy of order $\epsilon$, can be obtained in time
$$O(\kappa(A)^2 \emph{poly} (\log n)\|A\|_F/\epsilon),$$
where $\kappa(A)$ is the condition number of $A$.
\el

Another important fact associated with the complexity analysis of solving the preconditioned linear system \eqref{eqn:precondLin} is that during the quantum procedure, some quantum state can only be approximately obtained by using the SVE of $C$.
To check the accuracy of the generated state $|\phi\rangle$, we need to compare it with the exact one $|\psi\rangle$.

\bl \label{error-analysis}
Assume that
\[
|\phi\rangle=\frac{1}{\sqrt{Z}} \sum_{j=0}^{n-1} a_j \u_j, \hspace{.5cm}
|\psi\rangle=\frac{1}{\sqrt{W}} \sum_{j=0}^{n-1} b_j \v_j,
\]
where $\{ \u_j: j=0,\ldots,n-1\}$ and $\{ \v_j: j=0,\ldots,n-1\}$ are orthogonal basses, not necessarily to be unit. We assume that $|a_j-b_j|\leq \eta_0$, $\| \u_j - \v_j \|^2 \leq \eta_1$ for all $j$,  $|Z-W|\leq \eta_2$,  $\max_j\| \v_j \|^2=\eta_3$, and $1/\min_j\| \u_j \|^2 =\eta_4$. Then the error estimate reads
\begin{equation} \label{error1}
\| |\phi\rangle-|\psi\rangle \|^2 \leq 3\eta_1 \eta_4 +\frac{3\eta_2^2\eta_3 \eta_4}{W(\sqrt{W}+\sqrt{W-\eta_2})^2}+\frac{3n\eta_0^2 \eta_3}{W} .
\end{equation}
\el

The estimate of the error bound between $|\phi\rangle$ and $|\psi\rangle$ can be derived as follows.
\begin{eqnarray*}
\| |\phi\rangle-|\psi\rangle \|^2 & = & \ds\frac{1}{ZW} \sum_{j=0}^{n-1} \| \sqrt{W}a_j \textbf{u}_j - \sqrt{Z} b_j \textbf{v}_j \|^2 \\ \vspace{.2cm}
&\leq& \ds\frac{3}{ZW} \sum_{j=0}^{n-1} \Big( W|a_j|^2 \| \textbf{u}_j - \textbf{v}_j \|^2+|\sqrt{W}-\sqrt{Z}|^2 |a_j|^2 \| \textbf{v}_j \|^2 + Z |a_j-b_j|^2  \| \textbf{v}_j \|^2 \Big) \\\vspace{.2cm}
&\leq& \ds\frac{3}{ZW} \sum_{j=0}^{n-1} \Big( W |a_j|^2\eta_1 +|\sqrt{W}-\sqrt{Z} |^2 |a_j|^2\eta_3+Z\eta_0^2 \eta_3 \Big) \\\vspace{.2cm}
&=& 3\eta_1 \frac{1}{Z} \sum_{j=0}^{n-1} |a_j|^2 + \frac{3|W-Z|^2\eta_3}{W(\sqrt{W}+\sqrt{Z})^2} \frac{1}{Z} \sum_{j=0}^{n-1} |a_j|^2  +\frac{3n\eta_0^2 \eta_3}{W}  .
\end{eqnarray*}
Using the facts that $\sum_j |a_j|^2 / Z \leq \eta_4$, $|W-Z| \leq \eta_2$ and $\sqrt{W-\eta_2} \leq \sqrt{Z}$, we then obtain the estimation \eqref{error1}.

\subsection{Preconditioned linear solver}
\label{Preconditioned linear solver}

To design the quantum linear solver of the linear system   \eqref{eqn:precondLin}, we want the SVE of $C^{-1} A$. Such SVE demands the quantum states of columns of $C^{-1}A$ and $C^{-1}A$ itself, which further needs the SVE of $C$.

We first consider the construction of the preconditioner $C$ in quantum state.
Since $C = F^\dag \Lambda F$ and $F$ is Fourier transformation, we just need to focus on the diagonal matrix $\Lambda$.  The eigenvalues of $C$ or the diagonals of $\Lambda$ can be expressed by
\be \label{eigenvalues}
\lambda_k=\frac{1}{n} \sum_{p,q} \omega^{(p-q)k} A_{p,q}.
\ee

In the following we will form the state $|\lambda\rangle  = \tfrac{1}{\|C\|_F} \sum_{k=0}^{n-1}  \lambda_k |k\rangle $, where $\|C\|_F = (\sum_{k=0}^{n-1}  \lambda_k^2)^{\tfrac{1}{2}} $.
From the quantum state of $|A\rangle$, we can get
\[\ba{lll}\vspace{.2cm}
\ds\frac{1}{\|A\|_F} \sum_{p,q=0}^{n-1} A_{p,q}|p,q\rangle
&\mapsto& \ds\frac{1}{n\|A\|_F} \sum_{p,q,u,v=0}^{n-1} A_{p,q} \omega^{p  u - q  v}|u,v\rangle|u-v\rangle \\\vspace{.2cm}
&=& \ds\frac{1}{n\|A\|_F}   \sum_{p,q,k=0}^{n-1} A_{p,q} \omega^{(p -q)k} |k,k\rangle |0\rangle  +|0\rangle^\bot \\\vspace{.2cm}
&=& \ds \ds\frac{1}{\|A\|_F} \sum_{k=0}^{n-1} \lambda_k |k,k\rangle |0\rangle  +|0\rangle^\bot \\
&\mapsto& \ds\frac{1}{\|A\|_F} \sum_{k=0}^{n-1} \lambda_k |k\rangle |0 \rangle |0\rangle  +|0,0\rangle^\bot .
\ea\]
The probability to get $|\lambda\rangle$ is $\|C\|_F/\|A\|_F$. Performing measurements, we can get the state $| \lambda \rangle$ in time
\begin{equation}\label{eqn:Complex4ConstructC}
O(\|A\|_F \textmd{poly}(\log n) /\|C\|_F)=\widetilde{O}(\|A\|_F/\|C\|_F) .
\end{equation}

Therefore, $U_{\mathcal{N}}$ for $\Lambda$ can be implemented in time $\widetilde{O}(\|A\|_F/\|C\|_F)$, while $U_{\mathcal{M}}$ for $\Lambda$ is trivial. Thus we have the SVE of $\Lambda$, equivalently the SVE of $C$. Note that $\Lambda$ is diagonal, the SVD of $\Lambda$ is completely trivial if we know its diagonals explicitly. However,  a direct calculation according to the formula (\ref{eigenvalues})  will cost at least $O(n^2)$ to get $\Lambda$. In the quantum procedure above, we use a different method to construct the quantum state of the diagonal of $\Lambda$ within the time complexity as given in \eqref{eqn:Complex4ConstructC}.

Next, we consider how to form the quantum state $| C^{-1} A \rangle$. The basic idea is computing the inverse of $C$ based on  its SVE. As shown in HHL algorithm, such a procedure depends on the condition number of $C$.
By lemma \ref{linear-system}, the quantum state $|C^{-1}A_j\rangle$ of the $j$-th column  of $C^{-1}A$, which is proportional to $C^{-1}|A_j\rangle$,  can be prepared in time
\be \label{com1}
\widetilde{O}(\kappa(C)^2\|C\|_F \|A\|_F /\|C\|_F \epsilon)=\widetilde{O}( \|A\|_F\kappa(C)^2/\epsilon) .
\ee
%


Note that the quantum state of $|C^{-1}A\rangle$ equals
\[
|C^{-1}A\rangle=\frac{1}{\|C^{-1}A\|_F} \sum_{j=0}^{n-1} \|(C^{-1}A)_j\| |C^{-1}A_j\rangle|j\rangle
=\frac{1}{\|C^{-1}A\|_F} \sum_{j=0}^{n-1} \|A_j\| \|C^{-1}|A_j\rangle\| |C^{-1}A_j\rangle |j\rangle.
\]
Due to the parallelism of quantum computer, $|C^{-1}A\rangle$ can be also obtained in time (\ref{com1}). The error of obtaining $|C^{-1}A_j\rangle$ is bounded by $\epsilon$, however, the error of $|C^{-1}A\rangle$ will be enlarged by the summation. So we should analyze this error.

To estimate the error in generating $|C^{-1}A\rangle$, we need to estimate the errors in states $ |C^{-1}A_j\rangle$, the norms $\|C^{-1}|A_j\rangle\|$ and $\|C^{-1}A\|_F$, respectively due to lemma \ref{error-analysis}.
If we set $|A_j\rangle=\sum_k \alpha_{jk}F^\dag |k\rangle$, then $C^{-1}|A_j\rangle=\sum_k \alpha_{jk} \lambda_k^{-1} F^\dag |k\rangle$.
In the construction of $ |C^{-1}A_j\rangle$ by lemma \ref{linear-system}, the error in $\lambda_k^{-1}$ is bounded by $\epsilon$, and so the error of $C^{-1}|A_j\rangle$ is bounded by $\epsilon$ either.
Due to $\|C^{-1}|A_j\rangle\|^2=\sum_k |\alpha_{jk} \lambda_k^{-1}|^2 $ and an $\epsilon$ approximation of $\lambda_k^{-1}$,  the error of $\|C^{-1}|A_j\rangle\|^2$ is bounded by $\epsilon^2$.
Finally, the error of $\|C^{-1}A\|_F^2=\sum_j\|A_j\|^2 \|C^{-1}|A_j\rangle\|^2$ is bounded by $\|A\|_F^2\epsilon^2$.

Applying lemma \ref{error-analysis} with the parameters  $\eta_0=0$, $\eta_1=\epsilon^2$,   $\eta_2=\|A\|_F^2\epsilon^2$, $\eta_3=1/\min_k |\lambda_k|^2$, $\eta_4 = \max_k |\lambda_k|^2$,
the error of obtaining $|C^{-1}A\rangle$ is bounded by
\be\label{bound}
3\epsilon^2 \max_k |\lambda_k|^2  +\frac{3\|A\|_F^4\epsilon^4 \kappa^2(C) }{\|C^{-1}A\|_F^2(\|C^{-1}A\|_F+\sqrt{\|C^{-1}A\|_F^2-\|A\|_F^2\epsilon^2})^2 }.
\ee

Since $\frac{\| A \|_F}{\|C\|} \leq \|C^{-1} A\|_F \leq \|C^{-1}\| \|A\|_F$, we have $\frac{1}{\|C\|} \leq \frac{\|C^{-1} A\|_F}{\|A\|_F} \leq \|C^{-1}\| $.
Now we set $\|C^{-1}A\|_F^2=\beta\|A\|_F^2$, where $1/\max_j|\lambda_j|^2 \leq \beta\leq 1/\min_j|\lambda_j|^2$, then (\ref{bound}) can be written as
\be\label{bound1}
3\epsilon^2  \max_k |\lambda_k|^2 + \frac{3\epsilon^4 \kappa^2(C) }{\beta^2(1+\sqrt{1-\epsilon^2/\beta})^2 }.
\ee

We can perform a suitable scaling such that the singular values $|\lambda_k|$ of $C$ is smaller than 1 and larger than $1/\kappa(C)$.
Then $1\leq \beta\leq \kappa(C)^2$. Hence (\ref{bound1}) can be further simplified into
\be\label{bound2}
3\epsilon^2+3\epsilon^4\kappa(C)^2/\beta^2  .
\ee
To keep the error above bounded by  size $\epsilon_0$, we should choose $\epsilon$ such that $\epsilon^4\kappa(C)^2=\epsilon_0^2\beta^2$,
i.e., $\epsilon=\sqrt{\epsilon_0\beta/\kappa(C)}$.
Then the complexity to get the quantum state of $C^{-1}A$ is
\[
\widetilde{O}( \|A\|_F\kappa(C)^{5/2}/\sqrt{\epsilon_0\beta})=\widetilde{O}(\kappa(C)^{5/2} \|A\|_F^2/\sqrt{\epsilon_0}\|C^{-1}A\|_F).
\]

Finally, the complexity of the quantum linear solver based on SVE is summarized as follows.

\bt \label{Thm:ComplexityC}
The quantum state of the solution of $ Ax= b$ by using the preconditioner $C$ to accuracy $\epsilon$ can be obtained in time
\be
\widetilde{O}( \kappa(C)^{5/2}\kappa(C^{-1}A)^2\|A\|_F^2/\epsilon^{3/2}).
\ee
\et

Generally, it is not easy to compare the complexity given in theorem \ref{Thm:ComplexityC} with HHL algorithm and its variants, as well as the quantum algorithm given in \cite{WossnigZhaoPrakash_PRL18}. The following table is a list of already known quantum algorithms to solve linear system.

{\renewcommand\arraystretch{1.7}
\begin{table}[h]
\centering
\begin{tabular}{c|c|c}
  \hline
  Quantum algorithms & Complexity & Requirements \\\hline
  HHL algorithm \cite{harrow}                       & $O(s(A)\kappa(A)^2[\textmd{poly}\log(ns(A)\kappa(A)/\epsilon)]/\epsilon)$ & sparse \\\
  Ambainis' improved HHL  \cite{Ambainis_arXiv10}   & $O(s(A)\kappa(A)[\textmd{poly}\log(ns(A)\kappa(A)/\epsilon)]/\epsilon^3)$ & sparse \\
  CKS algorithm \cite{ChildsKothariSomma_SICOMP17}  & $O(s(A)\kappa(A)[\textmd{poly}\log(ns(A)\kappa(A)/\epsilon)])$ & sparse \\
  WZP algorithm \cite{WossnigZhaoPrakash_PRL18}     & $O( (\textmd{poly}\log n) \kappa(A)^2\|A\|_F/\epsilon)$ & no \\
  CJS algorithm \cite{CladerJacobsSprouse_PRL13}    & $O(s(A)^7\kappa(MA)(\log n)/\epsilon)$ & existence of SPAI \\
  our algorithm (theorem \ref{Thm:ComplexityC})     & $O( (\textmd{poly}\log n) \kappa(C)^{5/2}\kappa(C^{-1}A)^2\|A\|_F^2/\epsilon^{3/2})$ & no \\\hline
\end{tabular}
\caption{Comparison of quantum algorithms to solve linear system $Ax=b$, where $s(A)$ is the sparsity of $A$ and $\kappa(A)$ is the condition number of $A$. The last two solve the preconditioned linear system. The matrices $M$ and $C$
are the SPAI and circulant preconditioner of $A$ respectively.
}
\end{table}}

\br
For a good preconditioner, it is reasonable to assume that $\kappa(C) \ll \kappa(A)$ and $\kappa(C^{-1} A) \ll \kappa(A)$. That is, $\kappa(C)$ and $\kappa(C^{-1} A)$ can be assumed to be of $O(1)$. Under these conditions, the complexity can be further simplified to $\widetilde{O}(  \|A\|_F^2/\epsilon^{3/2})$.
\er

\br
When $\kappa(C)$ is of order one, the first term in \eqref{bound2} will be the dominate term.
Then the time complexity of $C^{-1}A$ is
$ \widetilde{O}( \|A\|_F \kappa(C)^2 / \epsilon  ) $.
The time complexity for solving \eqref{eqn:precondLin} is
$$ \widetilde{O}( \kappa(C)^2 \kappa(C^{-1} A)^2 \|A\|_F \|C^{-1} A\|_F  / \epsilon^2  ) . $$
\er


The above method for the circulant preconditioner $C$  can actually be extended to general cases.
We consider a general preconditioner $M$. The preconditioned linear system reads $ M^{-1} A x = M^{-1} b $.
Assume that the matrices $A$ and $M$ are stored in quantum state, for example, via qRAM.
To construct the quantum linear solver, we need two SVEs: $M$ and $M^{-1} A$. Specifically speaking,  we need the SVE of $M^{-1} A$, which is achieved by the SVE of $M$, via the following four steps.

(1) The SVE of $M$, which has the time complexity $\widetilde{O}( 1/ \epsilon )$ by lemma \ref{sve}.

(2) Calculation of $M^{-1} |A_j\rangle$ to form the state $|M^{-1}A \rangle$. Since we do not need to construct the preconditoner $M$ in quantum state as we do for the preconditioner $C$, the complexity term $\|A\|_F / \|M\|_F$ associated precondtioner construction similar to that in \eqref{eqn:Complex4ConstructC}  disappears.  Similar to the analysis above, where $\lambda_k$ is understood as the singular value, by using theorem \ref{Thm:ComplexityC} and noticing the disappearance of the complexity term $\|A\|_F / \|M\|_F$ associated with precondtioner construction, we have the time complexity
$
\widetilde{O}( \kappa(M)^{5/2}\kappa(M^{-1}A)^2 \|M\|_F \|A\|_F/\epsilon^{3/2})
$.


(3) The SVE of $M^{-1} A$, which has the time complexity $\widetilde{O} ( 1/ \epsilon )$ by lemma \ref{sve}.

(4) Linear solver associated with $M^{-1} A$, which costs $\widetilde{O} (\kappa(M^{-1}A)^2 \|M^{-1}A\|_F / \epsilon )$ by lemma \ref{linear-system}.

\bt
Given the matrices $A$ and $M$ are stored in quantum state, then the total time complexity for solving $ M^{-1} A x = M^{-1} b $ to accuracy $\epsilon$ in quantum computer is
\be
\widetilde{O}( \kappa(M)^{5/2}\kappa(M^{-1}A)^4  \|A\|_F \|M\|_F \|M^{-1}A\|_F /\epsilon^{9/2}) .
\ee
\et

\section{Conclusion}

In this paper, we present a new quantum algorithm based on circulant preconditioning technique to solve general linear systems, especially the dense cases with large condition numbers.
The main technique we applied here is the modified version of SVE (lemma \ref{sve}). This modified SVE will be more suitable to deal with the cases where we are given quantum inputs, and will have many other applications.
However, the new quantum algorithm to solve linear system (theorem \ref{Thm:ComplexityC}) depends on the Frobenius norm of the input matrix. As proved in \cite{harrow}, unless BQP=PSPACE, the condition number in the time complexity of solving linear system can not removed, so for general case, we cannot expect that $O(\kappa(C)^{5/2}\kappa(C^{-1}A)^2\|A\|_F^2)$ is small of size $O(\textmd{poly}\log n)$ all the time.
But it still remains a problem that how to improve the dependence of the complexity on $\|A\|_F$, since the result of \cite{WossnigZhaoPrakash_PRL18} is linear in $\|A\|_F$. Also, as suggested by the work of Childs et al \cite{ChildsKothariSomma_SICOMP17}, it may possible to improve the dependence on precision $\epsilon$ to polynomial of $\log 1/\epsilon$.


\section*{Acknowledgement}

H. Xiang is supported by the Natural Science
Foundation of China under grants 11571265, 11471253 and NSFC-RGC No. 11661161017.
C. Shao is supported by NSFC Project 11671388 and CAS Project QYZDJ-SSW-SYS022.

\end{document}